\documentclass[pre,twocolumn,preprintnumbers,amsmath,amssymb,nofootinbib,floatfix]{revtex4}

\usepackage{graphicx,bm}

\makeatletter

\def\graphicscale{\twocolumn@sw{0.3}{0.4}}
\def\graphicthreescale{\twocolumn@sw{0.3}{0.4}}

\begin{document}

\title{Decoherence dynamics of qubits coupled to systems
  at quantum transitions}

\author{Ettore Vicari} 
\affiliation{Dipartimento di Fisica dell'Universit\`a di Pisa
        and INFN, Largo Pontecorvo 3, I-56127 Pisa, Italy}

\date{\today}

\begin{abstract}

We study the decoherence properties of a two-level (qubit) system
homogeneously coupled to an environmental many-body system at a
quantum transition, considering both continuous and first-order
quantum transitions.  In particular, we consider a $d$-dimensional
quantum Ising model as environment system. We study the dynamic of the
qubit decoherence along the global quantum evolution starting from
pure states of the qubit and the ground state of the environment
system.  This issue is discussed within dynamic finite-size scaling
frameworks.  We analyze the dynamic finite-size scaling of appropriate
qubit-decoherence functions. At continuous quantum transitions, they
develop power laws of the size of the environment system, with a
substantial enhancement of the growth rate of the qubit decoherence
with respect to the case the environment system is in normal
noncritical conditions.  The enhancement of the qubit decoherence
growth rate appears much larger at first-order quantum transitions,
leading to exponential laws when increasing the size of the
environment system.

\end{abstract}

\maketitle


\section{Introduction}
\label{intro}

Decoherence generally arises when a quantum system interacts with an
environmental many-body system $S$. This issue is crucially related to
the emergence of classical behaviors in quantum
systems~\cite{Zurek-03,JP-09}, quantum effects such as interference
and entanglement~\cite{AFOV-08,FSDGS-18}, and it is particularly
relevant for the efficiency of quantum information
protocols~\cite{NV-book}.  The decoherence dynamics has been
investigated in some paradigmatic models, such as two-level (qubit)
systems interacting with many-body systems, in particular the
so-called central spin models, see, e.g.,
Refs.~\cite{Zurek-82,NP-93,SKL-02,CPZ-05,QSLZS-06,RCGMF-07,CFP-07,
  CP-08,DQZ-11,NDD-12,MSD-12,SND-16}, where the qubit is globally, or
partially, coupled to the environmental system $S$.

A typical problem concerns the coherence loss of the qubit during the
entangled quantum evolution of the global system, starting from pure
states of the qubit and the ground state of $S$.  The decoherence rate
may significantly depend on the quantum phase of $S$, and, in
particular, whether $S$ develops critical behaviors arising from
quantum transitions. Indeed the response of many-body systems at
quantum transitions is generally amplified by {\em critical} quantum
fluctuations. At quantum transitions, small variations of the driving
parameter give rise to significant changes of the ground state and
low-excitation properties of many-body systems~\cite{Sachdev-book}.
At first-order quantum transitions the ground-state properties appear
discontinuous in the infinite-volume limit, generally arising from
level crossings.  Continuous quantum transitions show continuous
change of the ground state at the transition point, and correlation
functions develop a divergent length scale.

Environmental systems at quantum transitions may significantly drive
the dynamics of the qubit decoherence.  An enhanced quantum
decoherence has been put forward~\cite{QSLZS-06} in the case of
continuous quantum transitions.  In this paper we return to this
issue, providing a quantitative scaling framework to support the
enhancement of the growth rate of the quantum decoherence, and extend
the analysis to the case the environmental system is at a first-order
quantum transitions.

We consider a qubit homogeneously coupled to a $d$-dimensional
many-body system $S$ of size $L$ (or equivalently with $N\sim L^d$
degrees of freedom). In particular, as environmental systems we
consider the paradigmatic $d$-dimensional quantum Ising models, whose
quantum phase diagrams present both continuous and first-order quantum
transitions~\cite{Sachdev-book}.  The two-level qubit system is
equally coupled to all $L^d$ spins of $S$.  We consider the standard
out-of-equilibrium protocol in which the initial global state is a
product of pure states of the qubit and $S$.  We study the quantum
decoherence dynamics during the quantum evolution of the global
system, as measured by its density matrix, obtained tracing out the
$S$-states.

We investigate the quantum decoherence dynamics when the environmental
Ising system experiences a quantum transition.  The decoherence
properties are analyzed within dynamic finite-size scaling frameworks.
At both continuous and first-order quantum transitions, dynamic
finite-size scaling behaviors arise from the interplay among the
coupling of the qubit with $S$, the Hamiltonian parameters of $S$
close to the quantum transition, and the size $L$ of $S$.  We show
that the {\em critical} conditions of the environmental system at
quantum transitions give rise to a substantial enhancement of the
growth rate of the decoherence dynamics with respect to noncritical
systems.  In particular, the decoherence growth rate at continuous
quantum transitions turns out to be characterized by power laws
$L^\zeta$ of the size $L$, with exponents $\zeta$ that are larger than
that of the volume $(L^d)$ law, expected for systems in normal
conditions.  The rate enhancement of the qubit coherence loss is even
more substantial at first-order quantum transitions. Indeed the
corresponding dynamic finite-size scaling theory predicts an
exponentially large decoherence growth rate, related to the
exponentially suppressed difference of the lowest levels in
finite-size many-body systems at first-order quantum transition.

The paper is organized as follows.  In Sec.~\ref{gset} we present the
general setting of the out-of-equilibrium problem that we consider.
In Sec.~\ref{genscacqt} we discuss the decoherence properties when the
environmental system is critical at a continuous quantum transition,
within a dynamic finite-size scaling framework, and show the enhanced
growth rate of decoherence with respect to normal conditions.
Sec.~\ref{OFSSfo} extends this analysis to first-order quantum
transitions, showing that the decoherence growth-rate enhancement is
even more pronounced, leading to exponential laws.  Finally in
Sec.~\ref{conclu} we summarize and draw our conclusions.

\section{General setting of the problem}
\label{gset}

We consider a $d$-dimensional quantum many-body system $S$ of size
$L^d$ with Hamiltonian
\begin{equation}
H_S(v) = H_c + v P_v \,,
\label{hsdef}
\end{equation}
where $P_v$ is the spatial integral of local operators, and
$[H_c,P_v]\neq 0$ and the parameter $v$ drives the quantum transition
located at $v=0$.  Then we consider a further two-level system
globally coupled to the many-body system, by the Hamiltonian term
\begin{equation}
H_{q} = w \, \Sigma^{(3)} P_v\,,
\label{hqdef}
\end{equation}
where the Pauli operator $\Sigma^{(3)}$ is associated with the two
states $|\pm\rangle$ of the qubit, so that $\Sigma^{(3)} | \pm \rangle
= \pm | \pm \rangle$.  Therefore the global Hamiltonian reads
\begin{equation}
H_{qS}(v,w) = H_c + (v + w\, \Sigma^{(3)}) \,P_v\,.
\label{hlamu}
\end{equation}
We are interested in the quantum evolution of the global system
starting from the initial $t=0$ condition
\begin{equation}
|\Psi_{qS}(t=0)\rangle = | q_0 \rangle \otimes | G_v \rangle\,,
\label{psit0}
\end{equation}
where $|q_0\rangle$ is a generic pure state of the qubit,
\begin{equation}
|q_0 \rangle = c_+ |+\rangle  + c_- |-\rangle,\qquad 
|c_+|^2 + |c_-|^2 = 1\,,
\label{iqstate}
\end{equation} 
and $| G_v \rangle$ is the ground state of the system with Hamiltonian
$H_S(v)$.  Then the global wave function describing the quantum
evolution for $t>0$ must be solution of the Schr\"odinger equation
\begin{equation}
i {\partial 
\over \partial t} |\Psi_{qS}(t)\rangle 
= H_{qS}(v,w) 
|\Psi_{qS}(t)\rangle\,.
\label{sceq}
\end{equation}
It can be written as
\begin{equation}
|\Psi_{qS}(t)\rangle = c_+ | + \rangle \otimes | \phi_{v+w}(t)
\rangle + c_- | - \rangle \otimes | \phi_{v-w}(t) \rangle\,,
\label{psit}
\end{equation}
where 
\begin{equation}
| \phi_{v\pm w}(t)\rangle = 
e^{-iH_S(v\pm w) t} |G_v\rangle\,,
\label{phipm}
\end{equation}
i.e., they are solutions of the Schr\"odinger equations for the system
$S$ only,
\begin{equation}
i {\partial 
\over \partial t} |\phi_{v\pm w}(t)\rangle 
= H_{S}(v\pm w) 
|\phi_{v \pm w}(t)\rangle\,,
\label{sceqb}
\end{equation}
with $|\phi_{v\pm w}(t=0)\rangle = | G_v\rangle$.  Note that the
expectation value $\langle \Psi_{qS}(t) | \Sigma^{(3)} |
\Psi_{qS}(t)\rangle = |c_+|^2 - |c_-|^2$ does not change along the
quantum evolution, thus it is fixed by the initial condition of the
qubit.

The quantum decoherence behavior can be inferred from the qubit
density matrix,
\begin{equation}
\rho_q(t) = {\rm Tr}_S \, \rho_{qS}(t)\,,\quad 
\rho_{qS}(t)=|\Psi_{qS}(t)\rangle \langle\Psi_{qS}(t)|\,,
\label{rhoq}
\end{equation}
where ${\rm Tr}_S$ is the trace over the $S$-states.  The {\em purity}
of the qubit during its quantum evolution can be quantified by the
trace of the square density matrix $\rho_q$, i.e.,
\begin{eqnarray}
{\rm Tr}\, \rho_q(t)^2 = 1 - 2 |c_+|^2 |c_-|^2 F_D(t)\, ,
\label{purity}
\end{eqnarray}
where
\begin{eqnarray}
F_D(t) = 1 - | \langle \phi_{v-w}(t) | \phi_{v+w}(t)
\rangle |^2\,,
\label{Sdef}
\end{eqnarray}
and $0\le F_D(t)\le 1$. The function $F_D$ measures the quantum
decoherence, quantifying the departure from a pure state. Indeed
$F_D(t)=0$ implies that the qubit is in a pure state, while $F_D(t)=1$
indicates that the qubit is maximally entangled, corresponding to a
diagonal density matrix
\begin{equation}
\rho_q = {\rm diag}[|c_+|^2,|c_-|^2].
\label{maxentrho}
\end{equation}
Of course, the time evolution of the 
decoherence function $F_D(t)\equiv
F_D(w,v,L,t)$ depends on the parameters of the global system, i.e.,
$v$ that measures the distance of the many-body system from the
quantum transition, the coupling $w$ between the qubit and the system,
and the size $L$ of the system.

Note that the overlap 
\begin{equation}
L_D(t)\equiv|\langle \phi_{v-w}(t) | \phi_{v+w}(t)\rangle|
\label{lddef}
\end{equation} 
entering the definition of $F_D$ can be interpreted as the fidelity or
Loschmidt echo, see, e.g., Ref.~\cite{QSLZS-06}, of the $S$-states
associated with two different quench protocols involving the isolated
system $S$. For both of them the system $S$ starts from the ground
state of the Hamiltonian $H_S(v)$ as $t=0$; then one considers, and
compares using $L_D$, the quantum evolutions at the same $t$, arising
from the sudden change of the Hamiltonian parameter $v$ to $v-w$ and
to $v+w$.

Noting that
\begin{equation}
\langle \phi_{v-w}(t) | \phi_{v+w}(t) \rangle=
\langle G_v| e^{i H_S(v-w) t} \;
e^{-i H_S(v+w) t} |G_v\rangle\,,
\label{parmu}
\end{equation}
one can easily show that $F_D$ is an even function of $w$.  Therefore,
since $F_D(0,v,L,t)=0$, and assuming an analytical behavior around
$\mu=0$ (at finite $L$ and $t$), we expect
\begin{equation}
F_D(w,v,L,t)  = {w^2\over 2} C_D(v,L,t) + O(w^4)\,
\label{dtmu2}
\end{equation}
for small values of $w$.  Thus the growth rate of the 
decoherence in the limit
of small qubit-$S$ coupling $w$ is described by the growth-rate
function
\begin{equation}
C_D(v,L,t) = \partial^2 F_D /\partial w^2|_{w=0}\,,
\label{cddef}
\end{equation}
for a given value $v$ of $H_S$. It measures the sensitivity of the
coherence properties of the subsystems to the qubit-$S$ coupling $w$.

The above setting can be straightforwardly extended to $n$-level
systems coupled to an environmental many-body system. The scaling
arguments we will report in the paper can be extended as well.

We also note that the above considerations can be straightforwardly
extended to the case the initial qubit state is not pure, but a mixed
state, thus described by a nontrivial density matrix. Of course, the
calculations become more cumbersome; however the function $F_D$
maintains its crucial role to describe the coherence properties
during the evolution of the global system.

As concrete examples of environmental systems $S$, we consider the
paradigmatic $d$-dimensional quantum Ising model defined on a $L^d$
lattice,
\begin{equation}
H_{I} = - J \, \sum_{\langle {\bf x},{\bf y}\rangle} \sigma^{(3)}_{\bf x}
\sigma^{(3)}_{\bf y} - g\, \sum_{\bf x} \sigma^{(1)}_{\bf x}\,,
\label{hisdef}
\end{equation}
where $\sigma^{(k)}$ are the Pauli matrices, the first sum is over all
bonds connecting nearest-neighbor sites $\langle {\bf x},{\bf
  y}\rangle$, while the other sums are over the sites.  We assume
$\hslash=1$, $J=1$, the lattice spacing $a=1$, and $g>0$.  

At $g=g_c>0$ (for one-dimensional quantum Ising systems $g_c=1$), the
model undergoes a continuous quantum transition belonging to the
$(d+1)$-dimensional Ising universality
class~\cite{Sachdev-book,ZJ-book,PV-02}, separating a disordered phase
($g>g_c$) from an ordered ($g<g_c$) one.  For any $g<g_c$, the
presence of a longitudinal external field $v$ coupled to
\begin{equation}
P_\ell = - \sum_{\bf x} \sigma^{(3)}_{\bf x}
\label{hvisdef}
\end{equation}
drives first-order quantum transitions along the $v=0$ line.

Then we consider a two-level qubit system, described by the Pauli
operator $\Sigma^{(3)}$ globally coupled to the Ising system by the
Hamiltonian term
\begin{equation}
H_{q} = w \,\Sigma^{(3)} P_\ell\,.
\label{hqis}
\end{equation}
We are interested in the coherence properties of the qubit when the
system $S$ is a $d$-dimensional Ising model with Hamiltonian
\begin{equation}
H_S(v) = H_I + v P_\ell\,,
\label{hivdef}
\end{equation}
cf. Eqs.~(\ref{hisdef}) and (\ref{hvisdef}), and the qubit coupling is
described by $H_{q}$ given in Eq.~(\ref{hqis}).

In the following sections we show that, at both continuous and
first-order quantum transitions of the environmental Ising system $S$,
the interplay among the coupling with the qubit, the Hamiltonian
parameters, the size $L$, gives rise to dynamic scaling behaviors of
the decoherence function $F_D(w,v,L,t)$, and correspondingly of its
growth-rate function $C_D(v,L,t)$.  For this purpose we consider
dynamic finite-size scaling frameworks, which allows us to
characterized the decoherence dynamics at both continuous and
first-order quantum transitions.  We derive the general features of
the dynamic finite-size scaling of $F_D$ and $C_D$, evidencing the
differences between continuous and first-order quantum transitions.

\section{The decoherence dynamics with a critical environmental system}
\label{genscacqt}

The theory of finite-size scaling at quantum transitions is well
established, see,
e.g.,~Ref.~\cite{CPV-14,CNPV-14,Barber-83,Privman-90} and references
therein.  The continuous quantum transition of the Ising model
(\ref{hisdef}) is characterized by two relevant parameters, $r\equiv
g-g_c$ and $v$ (such that they vanish at the critical point), with
renormalization-group dimension $y_r$ and $y_h$, respectively.  The
relevant finite-size scaling variables are
\begin{equation}
\kappa_r = L^{y_r} r\,,\qquad \kappa_v= L^{y_h} v\,.
\label{karh}
\end{equation}
The finite-size scaling limit is obtained by taking $L\to\infty$
keeping $\kappa_r$ and $\kappa_v$ fixed.

The equilibrium critical exponents $y_r$ and $y_h$ are those of the
$(d+1)$-dimensional Ising universality
class~\cite{Sachdev-book,ZJ-book,PV-02}. Therefore, for
one-dimensional systems they are $y_r=1/\nu=1$ and $y_h =
(d+3-\eta)/2= (4-\eta)/2$ with $\eta=1/4$.  For two-dimensional models
the critical exponents are not known exactly, but there are very
accurate estimates, see, e.g.,
Refs.~\cite{GZ-98,CPRV-02,Hasenbusch-10,KPSV-16,KP-17}; in
particular~\cite{KPSV-16} $y_r=1/\nu$ with $\nu=0.629971(4)$ and $y_h
= (5-\eta)/2$ with $\eta=0.036298(2)$.
For three-dimensional systems they assume mean-field values, $y_r=2$
and $y_h=3$, apart from logarithms.  The temperature $T$ gives rise to
a relevant perturbation at continuous quantum transitions, associated
with the scaling variable $\tau=L^z T$ where $z=1$ (for any spatial
dimension) is the dynamic exponent, characterizing the behavior of the
energy differences of the lowest-energy states and, in particular, the
gap $\Delta\sim L^{-z}$.  In the following we assume $T=0$.

A generic observable $O$ in the finite-size scaling limit behaves as
\begin{eqnarray}
 O(r,v, L) \approx L^{y_o} \, {\cal O}(\kappa_r, \kappa_v)\,,
  \label{cqtequi}
\end{eqnarray}
where the exponent $y_o$ is the renormalization-group dimension
associated with $O$, and ${\cal O}$ is a universal finite-size scaling
function.  The approach to such an asymptotic behavior is
characterized by power-law corrections, typically controlled by
irrelevant perturbations at the corresponding fixed
point~\cite{CPV-14}.  The equilibrium finite-size scaling at quantum
transitions has been also extended to quantum-information
concepts~\cite{GPSZ-06,AFOV-08,Gu-10,BAB-17}, such as the ground-state
fidelity and its susceptibility, which measure the change of the
ground state when varying the Hamiltonian parameters around a quantum
transition~\cite{RV-18}.

Out-of-equilibrium time-dependent processes require also an
appropriate rescaling of the time $t$, encoded by the scaling variable
\begin{equation}
\theta = L^{-z} t \sim \Delta(L)\, t\,.
\label{thetadefcqt}
\end{equation}
For example, we may consider the dynamic behavior of an isolated
system after a quench associated with a sudden change of the parameter
$v$, from $v$ to $v + w$ at $t=0$ (keeping $g$ fixed), starting from
the ground state $|G_v\rangle$.  The resulting quantum evolution of
the state is
\begin{equation}
|\phi(t)\rangle = e^{-iH_{S}(v+w) t} |G_{v} \rangle\,.
\label{phitq}
\end{equation}
This problem can be studied within a dynamic finite-size scaling
framework~\cite{PRV-18b}.  The dynamic finite-size scaling limit is
defined as the infinite-volume $L\to\infty$ limit, keeping the scaling
variables $\theta$, $\kappa_r$, $\kappa_v$, and
\begin{equation}
\kappa_{w} = L^{y_h}\,w
\label{kappadeh}
\end{equation}
fixed.  Then a generic observable $O$ in the dynamic finite-size
scaling limit is expected to behave as~\cite{PRV-18b}
\begin{eqnarray}
 O(r,v,w, L,t) \approx 
  L^{y_o} \, {\cal O}(\kappa_r, \kappa_v,\kappa_{w}, \theta)\,,
  \label{cqteq}
\end{eqnarray}
where again $y_o$ is the renormalization-group dimension of $O$, and
${\cal O}$ is a dynamic finite-size scaling function.  The equilibrium
finite-size scaling behavior is recovered in the limit $w \to 0$.

An analogous dynamic finite-size scaling is developed by the Loschmidt
echo $L_e$ associated with quench protocols, when suddenly changing
the driving parameter from $v$ to $v+w$.  The Loschmidt echo, defined
as
\begin{equation}
L_e(w,v,L,t) = |\langle
G_{v} | e^{-iH_{S}(v+w) t} |G_{v} \rangle|\,,
\label{ptdef}
\end{equation}
quantifies the deviation of the post-quench state at time $t > 0$ from
the initial $t=0$ ground state $|G_{v}\rangle$ associated with the
Hamiltonian $H_S(v)$.  It is expected to approach the asymptotic
dynamic finite-size scaling~\cite{PRV-18b}
\begin{eqnarray}
L_e(r,w,v,L,t) \approx {\cal L}_e (\kappa_r,\kappa_w,
\kappa_v,\theta)\,.
\label{Lasca}
\end{eqnarray}
This has been confirmed by numerical calculations within the
one-dimensional quantum Ising model around its continuous quantum
transition at $g_c=1$~\cite{PRV-18b}.  We also mention that the
dynamic finite-size scaling framework has been exploited to study the
scaling properties of work fluctuations after quenches at quantum
transitions~\cite{NRV-18}.

In order to derive the dynamic finite-size scaling behavior of the
decoherence function $F_D$, cf. Eq.~(\ref{purity}), we exploit its
close relation with the Loschmidt echo $L_D$ defined in
Eq.~(\ref{lddef}), between quantum states of $S$, along the quantum
evolutions arising from two different quench protocols of the isolated
system $S$, starting from the same state $|G_v\rangle$,
cf. Eq.~(\ref{parmu}).  Therefore, we expect that $F_D$ develops a
dynamic finite-size scaling analogous to that of the Loschmidt echo in
Eq.~(\ref{ptdef}) associated with standard quench protocols, as
reported in Eq.~(\ref{Lasca}).

To begin with, we consider quenches at the critical point $g=g_c$,
corresponding to $r=0$, driven by the parameter $v$.  According to the
above scaling arguments, we expect that the decoherence function
$F_D$, cf. Eq.~(\ref{Sdef}), develops the asymptotic dynamic 
finite-size scaling 
\begin{eqnarray}
F_D(r=0,w,v,L,t) \approx {\cal F}_D(\kappa_w, \kappa_v,\theta)\,,
\label{decqtsca}
\end{eqnarray}
with
\begin{eqnarray}
{\cal F}_D(\kappa_w=0, \kappa_v,\theta)=0\,,
\quad {\cal F}_D(\kappa_w, \kappa_v,\theta=0)=0\,.  \label{zerocond}
\end{eqnarray}
Note that the above dynamic finite-size scaling requires that also the
coupling $w$ between the qubit and $S$ is sufficiently small, indeed
the dynamic finite-size scaling limit requires that $\kappa_w =
L^{y_h} w$ must be kept constant in the large-$L$ limit. We do not
expect universal finite-size scaling behaviors without such a
rescaling, i.e., for generic finite values of $w$.

Moreover, Eq.~(\ref{decqtsca}) implies that the decoherence
growth-rate function $C_D$, cf. Eqs.~(\ref{dtmu2}) and (\ref{cddef}),
behaves as
\begin{equation}
C_D(r=0,v,L,t)\approx L^{2y_h} {\cal C}_D(\kappa_v,\theta)\,.
\label{cfhlt}
\end{equation}
This scaling equation characterizes the amplified $O(L^{2y_h})$ rate
of departure from coherence of the qubit when the environment system
$S$ is at a continuous quantum transition.  Indeed, in the case of
systems out of criticality one generally expects $C_D\sim L^d$, and
\begin{equation}
2y_h=d+3-\eta>d\,.
\label{2yrel}
\end{equation}

We may also consider the more general case when the parameter
$r=g-g_c$ is not zero, but sufficiently small to keep the system
within the critical region. The effects of a nonvanishing parameter
$r$ can be taken into account by adding a further dependence on
$\kappa_r$, cf. Eq.~(\ref{karh}), in the scaling function ${\cal
  C}_D(\kappa_v,\theta)$, i.e., we expect
\begin{equation}
C_D(r,v,L,t)\approx L^{2y_h} {\cal C}_D(\kappa_r,\kappa_v,\theta)\,.
\label{cfhltr}
\end{equation}
The scaling behavior in the thermodynamic limit can be formally
obtained by considering the limit $L\to\infty$ keeping fixed the
scaling variables
\begin{eqnarray}
&&\rho_v = \kappa_v \kappa_r^{-y_h/y_r} \equiv \xi_r^{y_h}
  \,v\,,\label{rhov}\\ &&\theta_v = \theta \kappa_r^{z/y_r} \equiv
  \xi_r^{-z}\,t\,,
\label{thetav}
\end{eqnarray}
where $\xi_r \sim r^{-1/y_r}$ is related to the diverging length scale
when approaching the critical point $r=0$ in the thermodynamic
limit. Therefore simple manipulations of the finite-size scaling
Eq.~(\ref{cfhltr}) lead to the following scaling behavior in the
thermodynamic limit
\begin{equation}
C_D(r,v,L\to\infty,t)\approx \xi_r^{2y_h} {\cal
  C}_\infty(\rho_v,\theta_v)\,,
\label{cflinf}
\end{equation}
obtained by replacing the scaling variables $\kappa_r,\,\kappa_v$ and
$\theta$ with $\rho_v,\,\theta_v$ and $L/\xi_r$, and considering the
thermodynamic limit $L/\xi_r\to\infty$.

The asymptotic behavior described by the dynamic finite-size scaling
at continuous quantum transitions is expected to be universal, i.e.,
independent of the microscopic features of the system $S$.  Its main
features only depend on the universality class of the continuous
quantum transition of $S$ and the general properties of the coupling
between the qubit and the system $S$. In the case at hand the qubit is
coupled to the order parameter of the magnetic transition.  Note that
the dynamic finite-size scaling functions generally depend on the
boundary conditions and the geometry of the system, while the power
laws of the observables and the scaling variables remain unchanged.
The approach to the dynamic finite-size scaling is expected to be
generally characterized by power-law suppressed corrections, as it
generally occurs at continuous quantum transitions~\cite{CPV-14}.

We may also consider the case in which the qubit is homogeneously
coupled to the transverse spin operators, i.e., we replace $P_\ell$,
cf. Eq.~(\ref{hvisdef}), with $P_t=-\sum_{\bf x} \sigma^{(1)}_{\bf
  x}$, and the qubit-$S$ coupling (\ref{hqis}) with 
\begin{equation}
H_{q,t} = u \,\Sigma^{(1)} P_t.
\label{hqtu}
\end{equation}
For simplicity we assume that $S$ is initially prepared in the ground
state for $v=0$ and a given $r=g-g_c$.  Using scaling arguments
analogous to those leading to Eq.~(\ref{decqtsca}), we arrive at the
dynamic finite-size scaling 
\begin{eqnarray}
F_D(u,r,L,t) \approx {\cal F}_D(\kappa_u, \kappa_r,\theta)\,,
\label{decqtsca2}
\end{eqnarray}
with $\kappa_r$ defined in Eq.~(\ref{karh}), and $\kappa_u = L^{y_r}
u$. This also implies
\begin{equation}
C_D(r,L,t)\approx L^{2y_r} {\cal C}_D(\kappa_r,\theta)
\label{cfhlt2}
\end{equation}
for the corresponding decoherence growth-rate function.  Note again
the enhancement of the decoherence dynamics, because $2y_r>d$.  The
decoherence dynamics of this central spin model, with the qubit
homogeneously coupled to the transverse spin variables of a
one-dimension Ising model, was also considered in
Ref.~\cite{QSLZS-06}; the scaling behavior of its numerical results
appears consistent with the dynamic finite-size scaling prediction
(\ref{decqtsca2}).

\section{The qubit decoherence with environmental systems at first-order 
quantum transitions}
\label{OFSSfo}

In this section we extend the dynamic finite-size scaling of the
decoherence dynamics to the case the environmental system $S$ is at a
first-order quantum transition, i.e. along the line $g<g_c$ of the
phase diagram of the $d$-dimensional Ising models.  We again consider
the quantum evolution of the global system starting from pure states
of both the qubit and the environmental Ising system $S$.

As shown by earlier works~\cite{CNPV-14,CNPV-15,CPV-15,PRV-18c}, the
finite-size scaling behaviors of isolated many-body systems at
first-order quantum transitions significantly depend on the type of
boundary conditions, in particular whether they favor one of the
phases or they are neutral, giving rise to finite-size scaling
characterized by exponential or power-law behaviors.  In the following
we consider Ising systems with boundary conditions that do not favor
any of the two magnetized phases, such as periodic and open boundary
conditions, which generally lead to exponential finite-size scaling
laws.

The first-order quantum transition line for $g<g_c$ are related to the
level crossing of the two lowest states $| \uparrow \rangle$ and $|
\downarrow \rangle$ for $v=0$, such that $\langle \uparrow |
\sigma_{\bf x}^{(3)} | \uparrow \rangle = m_0$ and $\langle \downarrow
| \sigma_{\bf x}^{(3)} | \downarrow \rangle = -m_0$ (independently of
${\bf x}$) with $m_0>0$.  The degeneracy of these states at $v=0$ is
lifted by the longitudinal field $v$. Therefore, $v = 0$ is a
first-order quantum transition point, where the longitudinal
magnetization $M = L^{-d} \sum_{\bf x} M_{\bf x}$, with $M_{\bf
  x}\equiv \langle \sigma_{\bf x}^{(3)} \rangle$, becomes
discontinuous in the infinite-volume limit.  The first-order 
quantum transition
separates two different phases characterized by opposite values of the
magnetization $m_0$, i.e.
\begin{equation}
\lim_{v \to 0^\pm}
\lim_{L\to\infty} M = \pm m_0\,.
\label{m0def}
\end{equation}
For one-dimensional systems~\cite{Pfeuty-70} $m_0 = (1 - g^2)^{1/8}$.

In a finite system of size $L$, the two lowest states are
superpositions of two magnetized states $| + \rangle$ and $| -
\rangle$ such that $\langle \pm | \sigma_{\bf x}^{(3)} | \pm \rangle =
\pm \,m_0$ for all sites ${\bf x}$.  Due to tunneling effects, the
energy gap $\Delta$ at $v=0$ vanishes exponentially as $L$
increases,~\cite{PF-83,CNPV-14}
\begin{equation}
\Delta(L) \sim e^{-c L^d}\,,
\label{del}
\end{equation}
apart from powers of $L$.  In particular, the energy gap $\Delta(L)$
of the one-dimensional Ising system (\ref{hisdef}) for $g<1$ is
exponentially suppressed as~\cite{Pfeuty-70,CJ-87}
\begin{eqnarray}
  \Delta(L) =  & 2 \, (1-g^2) g^L \, [1+ O(g^{2L})]
\label{deltaobc}
\end{eqnarray}
for open boundary conditions, and
\begin{eqnarray}
\Delta(L) \approx & 2 \,(\pi L)^{-1/2} (1-g^2) \, g^L
 \label{deltapbc}
\end{eqnarray}
for periodic boundary conditions.  The differences $\Delta_i\equiv
E_i-E_0$ for the higher excited states $(i>1$) are finite for $L\to
\infty$.

The emergence of a dynamic finite-size scaling after a quench protocol
is also expected along the first-order quantum transition line for
$g<g_c$~\cite{PRV-18b}, associated with a sudden change of the
parameter $v$, from $v$ to $v + w$ at $t=0$, starting from the ground
state $|G_v\rangle$.  Extending to generic dimensions the arguments of
Refs.~\cite{PRV-18,PRV-18b}, we identify the following scaling
variables
\begin{equation}
\kappa_v = {2 m_0 v \,L^d \over \Delta(L)}\,,\quad
\kappa_w = {2 m_0 w \,L^d \over \Delta(L)}\,,\quad
\theta = t \, \Delta(L)\,.
\label{scavarfoqt}
\end{equation}
In particular, the scaling variables $\kappa_v$ and $\kappa_w$ are the
ratios between the energy associated with the corresponding
longitudinal-field perturbations, which are approximately given by
$2m_0 v L^d$ and $2m_0 wL^d$ respectively, and the energy difference
$\Delta(L)$ of the two lowest states at $v=0$.  Then, the expected
dynamic finite-size scaling of the magnetization is~\cite{PRV-18b}
\begin{equation}
M(w,v,L,t) =  
m_0 \, {\cal M}(\kappa_w,\kappa_v,\theta)\,.
\label{mcheckfoqt}
\end{equation}
This dynamic finite-size scaling is expected to hold for any
$g<g_c$. The scaling function ${\cal M}$ is independent of $g$, apart
from trivial normalizations of the arguments.  The dynamic finite-size
scaling at first-order quantum transitions has been numerically
confirmed in the case of the one-dimensional Ising
model~\cite{PRV-18b}.  The approach to the asymptotic dynamic
finite-size scaling is expected to be exponential when increasing the
size of the system.  An analogous dynamic finite-size scaling applies
to the Loschmidt echo defined as in Eq.~(\ref{ptdef}), we expect
$L_e(w,v,L,t) \approx {\cal L}_e (\kappa_w, \kappa_v,\theta)$, which
is formally identical to Eq.~(\ref{Lasca}).

Then, using the same arguments of the previous section, i.e., noting
that the decoherence function $F_D$ can be written in terms of
quench-like amplitudes related to the environmental system only, we
conjecture an analogous dynamic finite-size scaling for the
decoherence function
\begin{eqnarray}
F_D(w,v,L,t) \approx {\cal F}_D(\kappa_w, \kappa_v,\theta)\,.
\label{decqtscafo}
\end{eqnarray}
Correspondingly, matching the expansion of the $F_D$ in powers of $w$
and that of ${\cal F}_D$ in powers of $\kappa_w$, we obtain the 
decoherence
growth-rate function
\begin{equation}
C_D(v,L,t) \approx {4 m_0^2 L^{2d}\over \Delta(L)^2}\,{\cal
  C}_D(\kappa_v,\theta)\,.
\label{cfhltfo}
\end{equation}
Therefore, when the environment system $S$ is at a first-order quantum
transition, the decoherence growth rate gets significantly enhanced,
increasing exponentially with $L$. Indeed the prefactor of
Eq.~(\ref{cfhltfo}) behaves as
\begin{equation}
{4 m_0^2 L^{2d}\over \Delta(L)^2} \sim \exp(b L^{d})\,,
\label{cdeppref}
\end{equation}
apart from  powers of $L$.

In the case of the quantum Ising systems with periodic or open
boundary conditions, the dynamic finite-size scaling functions can be
exactly computed, exploiting a two-level truncation of the
spectrum~\cite{CNPV-14,PRV-18}.  As shown in Ref.~\cite{PRV-18}, in
the long-time limit and for large systems, the scaling properties in a
small interval around $v=0$, more precisely for $m_0 |v|\ll
\Delta_2=O(1)$, are captured by a two-level truncation, which only
takes into account the two nearly-degenerate lowest-energy states.
The effective evolution is determined by the Schr\"odinger
equation~\cite{PRV-18}
\begin{equation}
i {d\over dt} \Psi(t) = H_{2}(v) \Psi(t) \,,
\label{sceq2l}
\end{equation}
where $\Psi(t)$ is a two-component wave function, whose components
correspond to the states $|+ \rangle$ and $|-\rangle$, and
\begin{eqnarray}
&&H_{2}(v) = - \beta \, \sigma^{(3)} + \delta \,
\sigma^{(1)}\, , \label{hrtds}\\
&&\beta = m_0 v L^d,\quad \delta = {\Delta(L)\over 2},\quad 
\kappa_v={\beta\over \delta}\,,\nonumber
\end{eqnarray}
where $\sigma^{(k)}$ are the Pauli matrices.  The initial condition is
given by the ground state of $H_2(v)$, i.e., by
\begin{equation}
|\Psi(w,v,L,t=0)\rangle = \sin(\alpha_v/2) \, |-\rangle -
\cos(\alpha_v/2) \, |+\rangle\,, \label{eigstatela0}
\end{equation}
with $\tan \alpha_v = \kappa_v^{-1}$ and $\alpha_v \in (0,\pi)$.  The
quantum evolution after quenching from $v$ to $v+w$ can be easily
obtained by diagonalizing $H_{2}(v+w)$, whose eigenstates are
\begin{eqnarray}
&& |0\rangle = \sin(\alpha_{v+w}/2) \, |-\rangle -  
\cos(\alpha_{v+w}/2) \, |+\rangle\,, \label{eigstate0la}\\
&& |1\rangle =  \cos(\alpha_{v+w}/2) |-\rangle +
\sin(\alpha_{v+w}/2) \, |+\rangle\,, \label{eigstate1la}
\end{eqnarray}
where $\tan \alpha_{v+w} = (\kappa_v+\kappa_w)^{-1}$ with
$\alpha_{v+w} \in (0,\pi)$. Their eigenvalue difference is given by
\begin{equation}
 E_1 - E_0 = \Delta(L) \; 
\sqrt{1 +  (\kappa_v+\kappa_w)^2}\,.
\end{equation}
Then, apart from an irrelevant phase, the time-dependent state evolves as
\begin{eqnarray}
&& |\Psi(w,v,L,t)\rangle =  
\cos\left({\alpha_v-\alpha_{v+w}\over 2}\right) |0\rangle  + \label{psitfo}\\
&&\quad +  
e^{-i \theta \sqrt{1 +  (\kappa_v+\kappa_w)^2}} 
\sin\left({\alpha_v-\alpha_{v+w}\over 2}\right) |1\rangle\,.
\nonumber
\end{eqnarray}
Note that the time-dependent wave function is written in terms of
scaling variables only.  The dynamic finite-size scaling of the
magnetization can be easily obtained~\cite{PRV-18b} by computing the
expectation value of the operator $\sigma^{(3)}$ over the state
$|\Psi(w,v,L,t)\rangle $.

\begin{figure}[tbp]
\includegraphics*[scale=\graphicscale]{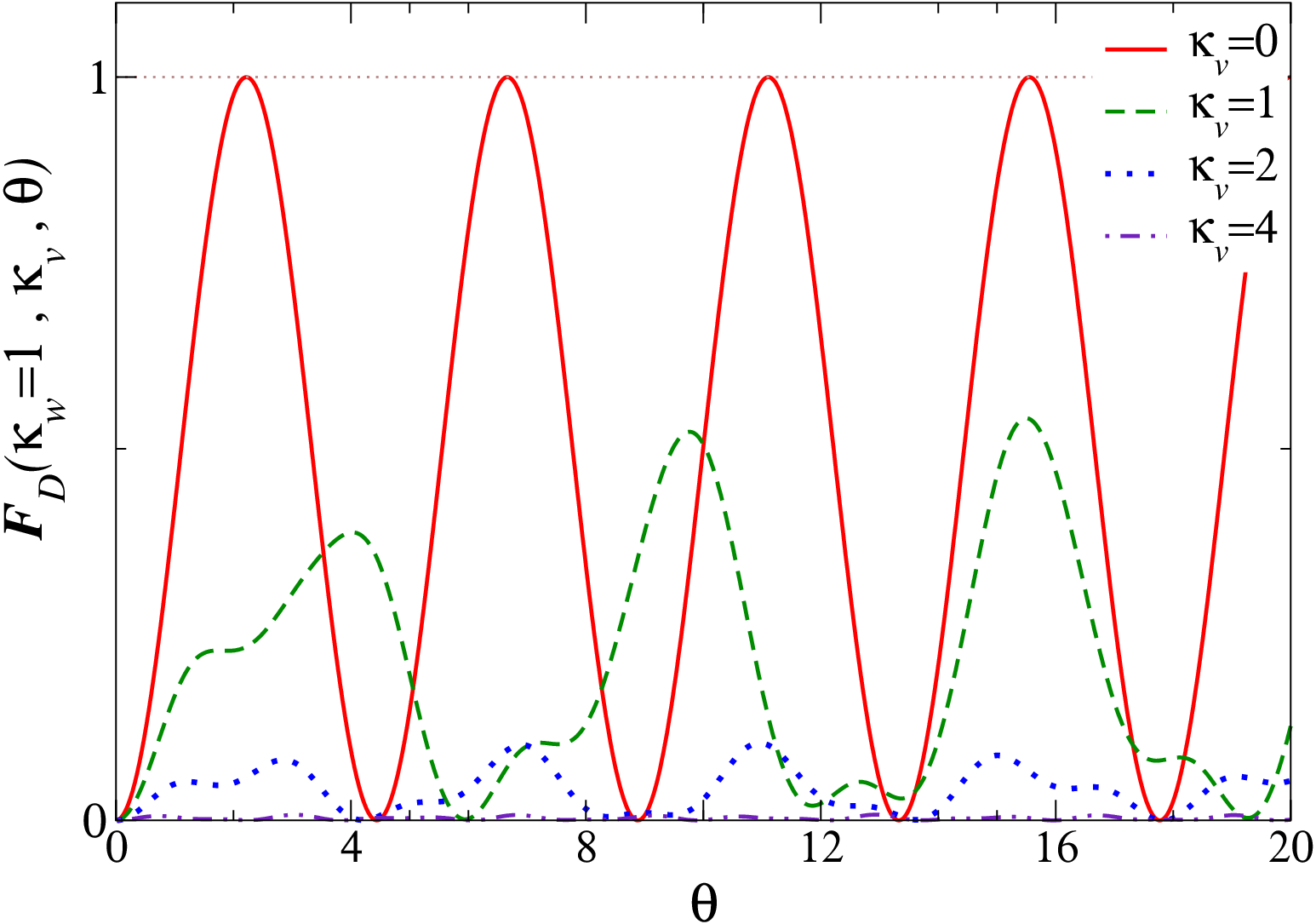}
\includegraphics*[scale=\graphicscale]{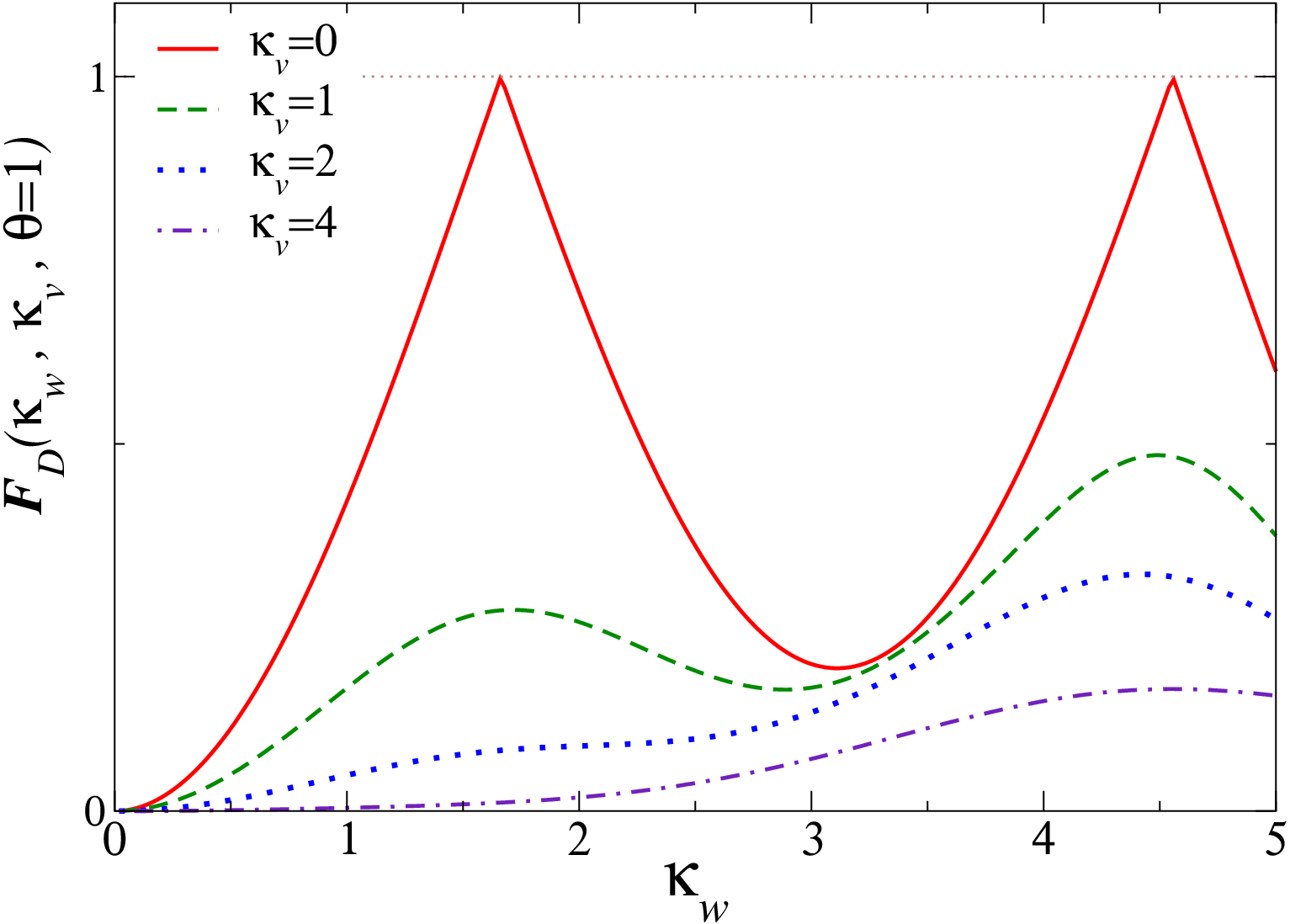}
\includegraphics*[scale=\graphicscale]{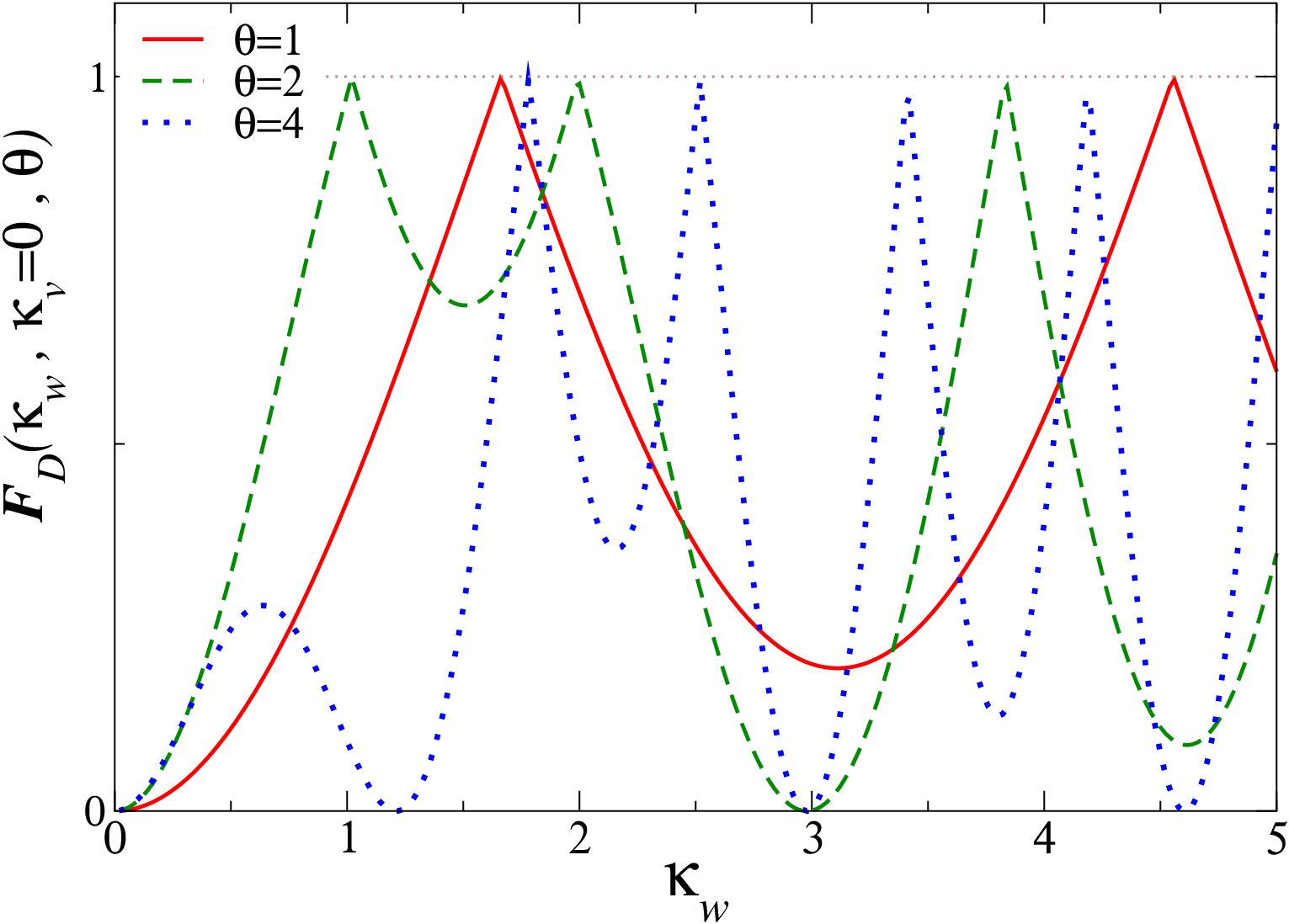}
\caption{Some plots of the scaling function ${\cal
    F}_D(\kappa_w,\kappa_v,\theta)$ associated with the 
decoherence dynamics at
  first-order quantum transitions, cf. Eq.~(\ref{decqtscafo}).  The
  top figure shows plots versus $\theta$ for $\kappa_w=1$ and some
  values of $\kappa_v$; the middle figure shows plots versus
  $\kappa_w$ for $\theta=1$ and some values of $\kappa_v$; the bottom
  figure shows plots versus $\kappa_w$ for $\kappa_v=0$ and some
  values of $\theta$. The units of the scaling variables can be easily
  inferred by taking into account that we set $\hslash=1$, $J=1$ and
  the lattice spacing $a=1$.}
\label{qubitD}
\end{figure}

The decoherence function $F_D$ can be straightforwardly obtained by
computing
\begin{equation}
F_D(w,v,L,t) = 
1 - |\langle \Psi(-w,v,L,t)|\Psi(w,v,L,t)\rangle|^2\,.
\label{amp}
\end{equation}
Using Eq.~(\ref{psitfo}), one can immediately see that $F_D(w,v,L,t)$
is a function of $\kappa_w$, $\kappa_v$ and $\theta$ only, confirming
the dynamic finite-size scaling Eq.~(\ref{decqtscafo}).  The resulting
expression is quite cumbersome, some plots are shown in
Fig.~\ref{qubitD}. The curves are also characterized by revivals,
typical of two-level systems.

The dynamic finite-size scaling of the decoherence growth-rate
function $C_D$ is obtained by computing
\begin{equation}
C_D(v,L,t) = {\partial^2 F_D\over \partial w^2}|_{w=0} =
\left( {\partial \kappa_w\over \partial w} \right)^2 
{\cal C}_D(\kappa_v,\theta)\,,
\label{cdtwol}
\end{equation}
which leads to the analytical result
\begin{equation}
{\cal C}_D(\kappa_v,\theta) = { 2[1 - {\rm
      cos}(\theta\sqrt{1+\kappa_v^2})]\over (1+\kappa_v^2)^2}\,.
\label{cdanres}
\end{equation}
Note the simple result for $\kappa_v=0$, 
\begin{equation}
{\cal C}_D(0,\theta)=2(1-\cos\theta),
\label{cdtheta0}
\end{equation}
 and that ${\cal C}_D(\kappa_v,\theta)$ vanishes for
 $\kappa_v\to\infty$.  We stress that the above dynamic finite-size
 scaling functions are expected to be independent of $g<g_c$ along the
 first-order transition line, apart from trivial $g$-dependent
 normalizations of the scaling variables.

We finally mention that a notable feature of one-dimensional quantum
Ising systems at first-order quantum transitions, with neutral
boundary conditions such as periodic and open boundary conditions, is
their rigidity with respect to external
perturbations~\cite{CNPV-14,PRV-18}, i.e.,~their response to global or
local longitudinal perturbations is analogous. Therefore, an analogous
quantum decoherence
 dynamics at the first-order transition line is expected in the case
of a local coupling between the longitudinal parameter $v$, the qubit
and the Ising chain, for example when replacing $P_\ell$, cf
Eqs.~(\ref{hvisdef}) and (\ref{hqis}), with
\begin{equation}
p_{\ell} = - \sigma^{(3)}_{x_c},\qquad
h_{q} = w \,\Sigma^{(3)} p_\ell
\label{hqisl}
\end{equation}
respectively, where $x_c$ is one of the sites of the chain
(sufficiently far from the boundaries).  The only difference is that
the relevant scaling variables turn into $\kappa_v = 2m_0 v/\Delta(L)$
and $\kappa_w = 2m_0 w/\Delta(L)$, instead of those reported in
Eq.~(\ref{scavarfoqt}).  They give rise to a two-level scenario as
well, with the same dynamic finite-size scaling functions.

\section{Conclusions}
\label{conclu}

We have investigated the decoherence dynamics of a two-level qubit
system globally and homogeneously coupled to a many-body spin system
$S$, such as a $d$-dimensional quantum Ising system, at a quantum
transition.  In particular, we have considered the out-of-equilibrium
quantum evolution of the global system starting from pure states of
both the qubit and $S$.  The decoherence dynamics of the qubit is
described by the time evolution of its density matrix, obtained
tracing out the states of $S$.  Its behavior can be characterized by
the decoherence function $F_D$ defined in Eq.~(\ref{purity}), which
quantifies the departure of the qubit from a pure state, independently
of its initial pure state. The sensitivity to the qubit-$S$ coupling
$w$ is measured by the decoherence growth-rate function
$C_D=\partial^2F_D/\partial w^2|_{w=0}$, cf. Eq.~(\ref{dtmu2}).

We have shown that the rate of the quantum decoherence gets enhanced
when the environmental system $S$ experiences a quantum transition.
At both continuous and first-order quantum transitions of $S$, the
interplay among the coupling between the qubit and $S$, the
Hamiltonian parameters and the size of $S$, during the quantum
evolution gives rise to scaling behaviors of the decoherence function
$F_D(w,v,L,t)$, cf. Eq.~(\ref{Sdef}), and the corresponding
decoherence growth-rate function $C_D(v,L,t)$, cf. Eq.~(\ref{cddef}),
in the limit of large size $L$ of $S$.  This is shown within dynamic
finite-size scaling frameworks, which allow us to determine the
behaviors of the decoherence functions at both continuous and
first-order quantum transitions of the environmental system $S$, in
appropriate dynamic finite-size scaling  limits.

We derive the general properties of the dynamic finite-size scaling of
the decoherence functions $F_D$ and $C_D$, evidencing the differences
between continuous and first-order quantum transitions. We show that
they are characterized by power laws of the size $L$ at continuous
quantum transitions, while exponential laws generally emerge at
first-order quantum transitions.  These behaviors represent a
substantial enhancement of the rate of the decoherence dynamics.  For
example, at continuous quantum transitions, when the qubit couples
longitudinally to the Ising model, the rate function turns out to
increase as $C_D \sim L^{\zeta}$ where $\zeta=2y_h = 15/8$ for $d=1$,
$\zeta=2y_h \approx 4.96$ for $d=2$, and $\zeta=2y_h=6$ for $d=3$
(apart from logarithms). Therefore they show a significant enhancement
of the decoherence growth rate, when compared with the general volume
$L^d$ law expected for systems in normal conditions. The decoherence
growth-rate enhancement appears even more substantial at first-order
quantum transitions, where $C_D\sim \exp(b L^{d})$ increases
exponentially.

Note that the main features of the dynamic finite-size scaling , such
as the general size dependence and the scaling functions, are expected
to be universal, i.e., they are expected not to depend on the
microscopic details of the models. Therefore, their predictions can be
extended to all continuous quantum transitions belonging to the same
Ising universality class with analogous coupling between qubit and
system.  An analogous statement holds for the dynamic finite-size
scaling with environmental systems at first-order quantum transitions.
In particular the dynamic finite-size scaling with environmental Ising
systems is expected to be the same, apart from normalizations, along
the first-order transition line for $g<g_c$, and in any system sharing
the same global properties, such as first-order quantum transitions
arising from an avoided two-level crossing phenomenon in the large-$L$
limit.

Finally we would like to stress that the dynamic finite-size scaling
frameworks, exploited to study the decoherence dynamics of qubit
coupled longitudinally and transversally to Ising systems at quantum
transitions, can be straightforwardly extended to general continuous
and first-order quantum transitions, and generic couplings of the
qubit to its environmental system.

\begin{acknowledgements}         
We acknowledge useful and interesting discussions with Davide Rossini.
\end{acknowledgements}

\end{document}